\documentstyle[12pt]{article}
\title{\bf On Duru-Kleinert Path Integral\\
in\\
Quantum Cosmology}
\vspace{20mm}
\author{M. A. Jafarizadeh$^{a,d}$ \thanks{e-mail: tabriz\_u@vax.ipm.ac.ir} ,
F. Darabi$^{b,c}$ \thanks{e-mail: f-darabi@cc.sbu.ac.ir} ,
A. Rastegar$^{a,c,d}$ \thanks{e-mail: rastgar@ark.tabrizu.ac.ir}\\
\\
\\
$^a${\small Department of Physics, Tabriz University, Tabriz, 51664, Iran.} \\
$^b${\small Department of Physics, Shahid Beheshti University, Tehran 19839, Iran.} \\
$^c${\small Department of Physics, Tarbiyat Moallem University, Tabriz, P.O.Box 51745-406, Iran.} \\
$^d${\small Institute for Studies in Theoretical Physics and Mathematics, Tehran, 19395-1795, Iran.}}
\begin{document}
\maketitle
\vspace{10mm}
\begin{abstract}
We show that the Duru-Kleinert fixed energy amplitude leads to the path integral
for the propagation amplitude in the closed FRW quantum cosmology with
scale factor as one degree of freedom. Then, using the Duru-Kleinert equivalence
of corresponding actions, we calculate the tunneling rate, with exact prefactor,
through the dilute-instanton approximation to first order in $\hbar$.
\end{abstract}
\vspace{50mm}
\section{Introduction}
$\; \; \; \;$ Tunneling is occurrent in almost all branches of physics including cosmology.
Some of the most significant examples are, the decay of the QFT false vacuum
in the inflationary model of universe \cite{B}, fluctuations as changes in the
topology in semiclassical gravity via tunneling process \cite{K},
tunneling rates for the production of pairs of black holes \cite{G},
minisuperspace quantum cosmology \cite{H}, and etc.

Here in this article we are concerned with the calculation of the tunneling
rate of the universe from nothing to the FRW universe based on the minisuperspace
model with only one degree of freedom by applying the dilute-instanton approximation
on the Duru-Kleinert path integral.

The Duru-Kleinert path integral formula for fixed energy amplitude is an alternative
approach to handle systems with singular potentials \cite{DK}. The heart of
this viewpoint is based on the Duru-Kleinert equivalence of actions, leading to
the same fixed energy amplitude, by means of arbitrary regulating functions originating
from the local reparametrization invariance.
On the other hand, reparametrization invariance is a basic property of general
relativity \cite{H}, so it is expected that:

One can establish a relation between Duru-Kleinert path integral for fixed
zero energy amplitude and the standard path integral in quantum cosmology, and
by using the corresponding Duru-Kleinert equivalence of actions it is possible
to work with the action that contains the standard quadratic kinetic term
instead of non-standard one.

In this paper we have studied these two subjects in the context of a ``mini-superspace''
model with only one degree of freedom.
In section ${\bf 2}$, the Duru-Kleinert path integral formula and Duru-Kleinert
equivalence of corresponding actions is briefly reviewed. In section ${\bf 3}$,
the standard path integral in quantum cosmology and its relation to Duru-Kleinert
path integral for closed FRW cosmology with only one degree of freedom (the scale factor) is investigated.
This section ends by introducing an equivalent standard quadratic action for
this cosmology.
Finally in section ${\bf 4}$, the rate of tunneling from nothing to the FRW
universe is calculated through the dilute instanton approximation to first order
in $\hbar$ \cite{B}, where its prefactor is calculated by the heat kernel method \cite{C},
using the shape invariance symmetry \cite{D}.

\section{Duru-Kleinert equivalence}

$\; \; \; \;$ In this section we briefly review Ref. 1.
The fundamental object of path integration is the time displacement
amplitude or propagator of a system, $ (X_b \: t_b\: | \: X_a \: t_a) $.
For a system with a time independent Hamiltonian, the object
$ (X_b \: t_b \: | \: X_a \: t_a) $ supplied by a path integral is the causal
propagator
\begin{equation}
(X_b \: t_b \: | \: X_a \: t_a)=\theta(t_a-t_b)<X_b|\exp(-i\hat{H}(t_b-t_a)/\hbar)|X_a>.
\end{equation}
Fourier transforming the causal propagator in the time variable, we
obtain the fixed energy amplitude
\begin{equation}
(X_b \: | \: X_a \: )_E = \int_{t_a}^\infty dt_b e^{iE(t_b-t_a)/\hbar}
(X_b \: t_b\: | \: X_a \: t_a)
\end{equation}
This amplitude contains as much information on the system as the propagator
$(X_b \: t_b\: | \: X_a \: t_a)$, and its path integral form is as follows:
\begin{equation}
(X_b \: | \: X_a)_E = \int_{t_a}^{\infty} dt_b \int {\cal D}x(t) e^{i{\cal A}_E/\hbar}
\end{equation}
with the action
\begin{equation}
{\cal A}_E = \int_{t_a}^{t_b} dt [\frac{M}{2}\dot{x}^2(t)-V(x(t))+E]
\end{equation}
where $ \dot{x} $ denotes the derivatives with respect to $ t $ .
In Ref. 1, it has been shown that fixed energy amplitude (3) is equivalent
with the following fixed energy amplitude,
\begin{equation}
(X_b \: | \: X_a)_E = \int_{0}^{\infty} dS [f_r(x_b)f_l(x_a)\int {\cal D}x(s)
e^{i{\cal A}_{E}^{f}/\hbar}]
\end{equation}
with the action
\begin{equation}
{\cal A}_{E}^{f} = \int_{0}^{S} ds \{ \frac{M}{2f(x(s))}x'^2(s)-f(x(s))
[V(x(s))-E] \}
\end{equation}
where $ f_r $ and $ f_l $ are arbitrary regulating functions and $ x'$ denotes
the derivatives with respect to $s$.

The actions $ {\cal A}_E $ and $ {\cal A}_{E}^{f} $,
both of which lead to the same fixed-energy amplitude $ (X_b \: | \: X_a)_E $ are called
Duru-Kleinert equivalent \footnote{Of course a third action
$ {\cal A}_{E,\varepsilon}^{DK} $ is also Duru-Kleinert equivalent of
$ {\cal A}_E $ and $ {\cal A}_E^f $ which we do not consider here.}.
The motivation of Duru and Kleinert, using this equivalence, was to investigate
the path integrals for singular potentials.

In the following section we show that one can use this equivalence to
investigate the quantum cosmological models, with one degree of freedom.\, To
see this rewrite the action $ {\cal A}_{E}^{f} $ in a suitable form such that
it describes a system with zero energy; as only in this sense can we describe
a quantum cosmological model with zero energy.\\
Imposing $ E = 0 $ in (6), with a simple manipulation, gives
\begin{equation}
{\cal A}_{E}^{f} = \int_{0}^{1} ds' S f(X(s')) \{ \frac{M}{2[Sf(X(s'))]^2}
\dot{X}^2(s')-V(X(s')) \}
\end{equation}
where $ \dot{X} $ denotes the derivative with respect to new parameter $ s' $ defined by
\begin{equation}
s' = S^{-1} s
\end{equation}
with $S$ as a  dimensionless scale parameter.

After a Wick rotation $ s'=-i\tau $, we get the required Euclidean action and
the path integral
\begin{equation}
I_{0}^{f} = \int_{0}^{1} d\tau S f(X(\tau)) \{ \frac{M}{2[Sf(X(\tau))]^2}
\dot{X}^2(\tau)+V(X(\tau)) \}
\end{equation}
\begin{equation}
(X_b \: | \: X_a) = \int_{0}^{\infty} dS [f_r(X_b)f_l(X_a) \int{\cal D}X(\tau)
e^{{-I_{0}^{f}}/\hbar}].
\end{equation}
where $\tau$ is the Euclidean time. We will use eqs. (9) and (10) in the following section.

\section{Path integral in Quantum Cosmology }

$\; \; \; \;$ General formalism of quantum cosmology is based on the Hamiltonian formulation
of general relativity, specially Dirac quantization procedure in which the wave
function of the universe $ \Psi $ is obtained by solving the Wheeler-DeWitt
equation
\begin{equation}
\hat{H} \Psi = 0.
\end{equation}
A more general and more powerful tool for calculating the wave function is the path
integral. In ref. 2 it is shown that the path integral for the propagation
amplitude between fixed initial and final configurations can be written as
\begin{equation}
(X_b \: | \: X_a)=\int_{0}^{\infty} dN <X_b,N \: | \: X_a,0> \: =\int_{0}^{\infty}
dN \int {\cal D}X e^{-I[X(\tau),N]/\hbar}
\end{equation}
where $ <X_b,N \: | \: X_a,0> $ is a Green function for the Wheeler-DeWitt
equation and $N$ is the lapse function. The Euclidean action $I$ is defined on
minisuperspace in the gauge $ \dot{N} = 0 $ as
\begin{equation}
I[X(\tau),N] = \int_{0}^{1} d\tau N [\frac{1}{2N^2}f_{ab}(X)\dot{X}^{a}\dot{X}^{b}
+V(X)]
\end{equation}
where $ f_{ab}(X) $ is the metric defined on minisuperspace and has indefinite signature.

Here we consider a model in which the metric $ f_{ab}(X) $ is defined by only
one component and takes the following Euclidean action \cite{H},
\begin{equation}
I =  \int_{0}^{1} d\tau N [\frac{R\dot{R}^2}{2N^2}+\frac{1}{2}(R-\frac{R^3}
{a_{0}^2})]
\end{equation}
where $R$ is the scale factor and $ R_{0}^2 = \frac{3}{\Lambda} $ is interpreted
as the minimum radius of the universe after tunneling from nothing \cite{H} ($ \Lambda $
is cosmological constant).\, This model describes the closed FRW universe with
one degree of freedom $R$.\\
Now we rewrite the action (14) as
\begin{equation}
I = \int_{0}^{1} d\tau N R^{-1} [\frac{\dot{R}^2}{2N^2R^{-2}}+\frac{1}{2}
(R^2-\frac{R^4}{R_{0}^2})].
\end{equation}
Comparing this action with (9) (with $ M = 1 $) we find that by choosing
\begin{equation}
N R^{-1} = S f(R)
\end{equation}
we obtain (9) in the form
\begin{equation}
I = I_{0}^f = \int_{0}^{1} d\tau S f(R) [\frac{\dot{R}^2}{2[Sf(R)]^2}+V(R)]
\end{equation}
such that
\begin{equation}
V(R) = \frac{1}{2}(R^2-\frac{R^4}{R_0^2}).
\end{equation}
The gauge $ \dot{N} = 0 $ gives
\begin{equation}
f(R) = C R^{-1}
\end{equation}
where $C$ is a constant which we set to $C=R_a^{-1}$ so that
$$
S = N\,R_a.
$$
Now, one can show that the path integral (10) corresponds
to the path integral (12).\\
To see this, assume
\begin{equation}
f_r(R) = 1 \;\;\; , \;\;\; f_l(R) = f(R)
\end{equation}
so that the path integral (10) can be written as
\begin{equation}
(R_b \: | \: R_a) = \int_{0}^{\infty} dN \int {\cal D}R \:
e^{{-I_0^f}/\hbar}
\end{equation}
where $ I_0^f $ is given by (17). This shows that the Duru-Kleinert
path integral (21) is exactly in the form of (12) as a path integral
for this cosmological model. Now, using the Duru-Kleinert equivalence,
we can work with the standard quadratic action
\begin{equation}
I_0 = \int_{\tau_a}^{\tau_b} d\tau [\frac{1}{2}\dot{R}^2(\tau)+\frac{1}{2}
(R^2-\frac{R^4}{R_{0}^2})]
\end{equation}
instead of the action (17) or (14), where a Wick rotation
with $ E = 0 $ has also been used in the equation (4).

\section{Tunneling rate}

$\; \; \; \;$ The Euclidean type Lagrangian corresponding to the action (22) has
the following quadratic form
\begin{equation}
L_E = \frac{1}{2}\dot{R}^2 + \frac{1}{2}(R^2 - \frac{R^4}{R_0^2}).
\end{equation}
The corresponding Hamiltonian is obtained by a Legendre transformation
\begin{equation}
H_E = \frac{\dot{R}^2}{2} - \frac{1}{2}(R^2 - \frac{R^4}{R_0^2}).
\end{equation}
Imposing $ H_E = 0 $ gives a nontrivial ``instanton solution''as
\begin{equation}
R(\tau) = \frac{R_0}{\cosh(\tau)},
\end{equation}
which describes a particle rolling down from the top of a potential $ -V(R) $
at $ \tau \rightarrow -\infty $ and $ R = 0 $, bouncing back at $ \tau = 0 $ and
$ R = R_0 $ and finally reaching the top of the potential at $ \tau \rightarrow
+\infty $ and $ R = 0 $.\\
The region of the barrier $ 0 < R < R_0 $ is classically forbidden for the zero energy
particle, but quantum mechanically it can tunnel through it with a tunneling
probability which is calculated making use of the instanton solution (25).\\
The quantized FRW universe is mathematically equivalent to this particle, such
that the particle at $ R = 0 $ and $ R = R_0 $ represents ``nothing'' and ``FRW''
universes respectively. Therefore one can find the probability
$$
|<FRW(R_0) \: | \: nothing>|^2 .
$$
The rate of tunneling $ \Gamma $ is calculated through the dilute instanton
approximation to first order in $\hbar$ as \cite{B}
\begin{equation}
\Gamma = [\frac{det'(-\partial_{\tau}^2 + V''(R))}{det(-\partial_{\tau}^2 + \omega^2)}]^{-1/2}
e^{\frac{-I_0(R)}{\hbar}} [\frac{I_0(R)}{2\pi\hbar}]^{1/2}
\end{equation}
where det' is the determinant without the zero eigenvalue,\, $ V(R)'' $ is the
second derivative of the potential at the instanton solution (25 ), $ \omega $
corresponds to the real part of the energy of the false vacuum $ (|nothing>) $
and $ I_0(R) $ is the corresponding Euclidean action.
The determinant in the numerator is defined as
\begin{equation}
det'[-\partial_{\tau}^2 + V''(R)] \equiv \prod_{n=1}^{\infty}|\lambda_n|
\end{equation}
where $ \lambda_n $ are the non-zero eigenvalues of the operator
$ -\partial_{\tau}^2 + V''(R) $.\\
The explicit form of this operator is obtained as
\begin{equation}
O \equiv [-\frac{d^2}{dx^2} + 1 - \frac{6}{\cosh^2(x)}]
\end{equation}
where we have used Eqs.(18) and (25) and a change of variable $ x = \tau $
has been done.\\
Now, we can calculate the ratio of the determinants as follows:
First we explain very briefly how one can calculate the determinant of an
operator through the heat kernel method \cite{C}. We introduce the generalized
Riemann zeta function of the operator $A$ by
\begin{equation}
\zeta_A(s) = \sum_{m} \frac{1}{|\lambda_m|^s}
\end{equation}
where $ \lambda_m $ are eigenvalues of the operator $A$,\, and the determinant
of the operator $A$ is given by
\begin{equation}
det \, A = e^{-\zeta'_{A}(0)}.
\end{equation}
On the other hand $ \zeta_A(s) $ is the Mellin transformation of the heat kernel
$ G(x,\, y,\, t)$
\footnote{Here $t$ is a typical time parameter.}
which satisfies the following heat diffusion equation,
\begin{equation}
A \, G(x,\,y,\, t) = -\frac{\partial \, G(x,\,y,\, t)}{\partial t}
\end{equation}
with an initial condition $ G(x,\,y,\,0) = \delta(x - y) $.\, Note that
$ G(x,\,y,\, t) $ can be written in terms of its spectrum
\begin{equation}
G(x,\,y,\, t) =  \sum_{m} e^{-\lambda_{m}t} \psi_{m}^{*}(x) \psi_{m}(y).
\end{equation}
An integral is written for the sum if the spectrum is continuous.
From relations (30) and (31) it is clear that
\begin{equation}
\zeta_{A}(s) = \frac{1}{\Gamma(s)} \int_{0}^{\infty} dt \, t^{s-1}
\int_{-\infty}^{+\infty} dx \, G(x,\,x,\, t).
\end{equation}
Now, in order to calculate the ratio of the determinants in (26), called a
prefactor, we note that it is required to find the difference of the functions
$ G(x, y, t) $.\\
We rewrite the operator (28) as:
\begin{equation}
[ (-\frac{d^2}{dx^2}-\frac{2(2+1)}{\cosh^{2}(x)}+4)-3 ].
\end{equation}
This is the same as the operator which appears in Ref.5, for values of
$ l = 2 $, $ h = -3 $; so the heat kernel $ G(x, y, t)$ corresponding to
the operator (34) is given by
\begin{equation}
G_{\Delta_{2}(0)-3}(x, y, t) = \frac{e^{-(4-3)t}}{2\sqrt{\pi t}}
e^{-(x-y)^2/{4 t}}
\end{equation}
and
\begin{equation}
\begin{array}{lll}
G_{\Delta_{2}-3}(x, y, t) & = & \psi_{2,0}^{\ast}(x)\psi_{2,0}(y)e^{-|-3|t}\\
	& & \\
	& & +\int_{-\infty}^{+\infty} \frac{dk}{2\pi} \frac{e^{(1+k^2)t}}{(k^2+1)(k^2+4)}
[ \left( B_{2}^{\dag}(x) B_{1}^{\dag}(x) e^{-ikx} \right) ]
[ \left( B_{2}^{\dag}(y) B_{1}^{\dag}(y) e^{iky} \right) ].
\end{array}
\end{equation}
The functions $ \psi_{l,m} $ and $ \psi_{l,k} $ are the eigenfunctions corresponding
to discrete spectrum $ E_{l,m} = m(2l-m) $ and continuous spectrum $ E_{l,k} =
l^2 + k^2 $ of the following operator
$$
-\frac{d^2}{dx^2}-\frac{l(l+1)}{\cosh^2(x)}+l^2
$$
respectively, and are given by \cite{D}
$$
\psi_{l,m}(x) = \sqrt{\frac{2(2m-1)!}{\prod_{j=1}^{m} j(2l-j)}}\frac{1}{2^m(m-1)!}
B_{l}^{\dag}(x)B_{l-1}^{\dag}(x) \cdots B_{m+1}^{\dag}(x)\frac{1}{\cosh^{m}(x)}
$$
and
$$
\psi_{l,k}(x) = \frac{B_{l}^{\dag}(x)}{\sqrt{k^2+l^2}}
\frac{B_{l-1}^{\dag}(x)}{\sqrt{k^2+(l-1)^2}} \cdots
\frac{B_{1}^{\dag}(x)}{\sqrt{k^2+1^2}} \frac{e^{ikx}}{\sqrt{2 \pi}}
$$
where
$$
B_{l}(x) := \frac{d}{dx}+l \; \tanh(x), \hspace{10mm}
B_{l}^{\dag}(x) := - \frac{d}{dx}+l \; \tanh(x).
$$
Now, we can write
$$
\int_{-\infty}^{+\infty} dx [G_{\Delta_{2}-3}(x, x, t) - G_{\Delta_{2}(0)-3}(x, x, t)]
= e^{-3t} -\frac{3}{\pi} e^{-t} \int_{-\infty}^{+\infty} dk \: \frac{(k^2+2)e^{-k^2 t}}
{(k^2+1)(k^2+4)}
$$
and
$$
\begin{array}{lll}
\zeta_{\Delta_{2}-3}(s) - \zeta_{\Delta_{2}(0)-3}(s) & = &
\frac{1}{3^s} -\frac{3}{\pi} \int_{-\infty}^{+\infty} \frac{dk}{(k^2+1)^s(k^2+4)}
-\frac{3}{\pi} \int_{-\infty}^{+\infty} \frac{dk}{(k^2+1)^{s+1}(k^2+4)} \\
	& & \\
	& = & \frac{1}{3^s} -\frac{3}{\sqrt{\pi}} \frac{1}{2^{2s+1}} \frac{\Gamma(s+\frac{1}{2})}{\Gamma(s+1)}
	\{ F(s, s+\frac{1}{2}, s+1; \frac{3}{4}) \\
	& & \\
	& & \hspace{36mm} + \frac{2s+1}{8(s+1)} F(s+1, s+\frac{3}{2}, s+2; \frac{3}{4}) \}.
\end{array}
$$
Thus, we have
$$
\hspace{-40mm} \zeta_{\Delta_{2}-3}(0) - \zeta_{\Delta_{2}(0)-3}(0) = -1 \: , \hspace{10mm}
\zeta'_{\Delta_{2}-3}(0) - \zeta'_{\Delta_{2}(0)-3}(0) = Ln(12)
$$
So we get the following value for the prefactor in (26)
\begin{equation}
{\it prefactor} =  \: e^{-ln(12)} \: = \: \frac{1}{12} \:
\end{equation}
Substituting the result (37) in Eq. (26) gives the tunneling rate
of the universe from nothing to the FRW universe

$$
\Gamma = \frac{2}{\sqrt{\pi \hbar}} R_0 e^{- \frac{2 R_0^2}{3\hbar}} \: + \: O(\hbar).
$$
in a good agreement with the result obtained by WKB approximation (Atkatz \cite{H}).
\section{Conclusions}

$\; \; \; \;$ We have shown in this paper that one can obtain the path integral formula of
quantum cosmology by Duru-Kleinert path integral formula, at least for a model
with one degree of freedom. The prime point is that, to the extent the path integral
quantum cosmology is concerned, one can work with the standard action instead
of non standard one, by using the Duru-Kleinert equivalence of the actions.
This will be valuable in simplifying the possible technical problems which may appear in working with
non standard actions. We have concentrated on the model with only one degree of
freedom. Whether this procedure works for higher degrees of freedom is a question
which requires further investigation.

\end{document}